\documentclass{article}
\usepackage[centertags]{amsmath}

\begin{document}

\title{The classification of traveling wave solutions and superposition of
multi-solutions to Camassa-Holm equation with dispersion}

\author{ Chengshi Liu \\Department of Mathematics\\Daqing Petroleum Institute\\Daqing 163318, China
\\Email: chengshiliu-68@126.com}

 \maketitle

\begin{abstract}
Under the traveling wave transformation,  Camassa-Holm equation with
dispersion is reduced to an integrable ODE whose general solution
can be obtained using the trick of one-parameter group. Furthermore
combining complete discrimination system for polynomial, the
classifications of all single traveling wave solutions to  the
Camassa-Holm equation with dispersion is obtained.
 In particular, an affine subspace structure in the set of the
 solutions of the reduced ODE is
 obtained. More general, an implicit linear structure in Camassa-Holm equation with
dispersion is found. According to the linear structure, we obtain
 the superposition of multi-solutions to Camassa-Holm equation with
dispersion.


\end{abstract}

\section{Introduction}

Camassa-Holm equation
(CH)((\cite{CH},\cite{CHH},\cite{ACH},\cite{ACH2})reads
\begin{eqnarray}\label{CH}
u_t+2ku_x-u_{xxt}+3uu_x=2u_xu_{xx}+uu_{xxx},
\end{eqnarray}
which describes shallow water waves. CH equation has been studied
from several aspects([1-15]). Camassa and Holm found that this
equation exhibits "peakons" when $k=0$, that is, solitary wave
solutions with discontinuous first derivative at their crest. This
peakons have the following form
\begin{equation}
u=c\exp(-|x-ct|).
\end{equation}
For arbitrary value of $k$, Liu et al (\cite{lz1,lz2})show that the
CH equation has peakons as
\begin{equation}
u=(c+k)\exp(-|x-ct|)-k.
\end{equation}
Liu et al (\cite{lz3})also obtain a kind of generalized kink and
anti-kink wave solutions. In 2001, based on the CH equation, Dullin
et al (\cite{d1})drive Camassa-Holm equation with dispersion, its
form is as follows
\begin{eqnarray}\label{GCH}
u_t+2ku_x+3uu_x-\varepsilon(u_{xxt}+2u_xu_{xx}+uu_{xxx})+\gamma
u_{xxx}=0,
\end{eqnarray}
where $k, \varepsilon$ and $\gamma$ are constants. If $k=\epsilon=0$
and $\gamma=1$, then Camassa-Holm equation with dispersion becomes
KdV equation. If $\varepsilon=1$ and $\gamma=0$, Camassa-Holm
equation with dispersion  becomes CH equation . Dullin et
al(\cite{d1}) obtain peakons in the case of $k=\gamma=0$. Guo and
Liu, Zhang, Tang and Yang (\cite{qi,bo,guo,zh,ta})give some peakons
in some special cases. These peakons are so called weak solutions or
generalized solutions. Here
we consider classical solution.\\

In the present paper, we give the complete classification of
classical traveling wave solutions of Camassa-Holm equation with
dispersion. Applying the invariance property of the reduced equation
of Camassa-Holm equation with dispersion under a one-parameter
group, we obtain the general solution of its reduced equation.
Furthermore using the complete discrimination system for polynomial,
we  give the classifications of  traveling wave solutions to
Camassa-Holm equation with dispersion. This complete result is
valuable to study the physical properties of Camassa-Holm equation
with dispersion. In addition, we find a two dimensional affine
subspace structure in the set of the solutions of the reduced ODE.
More general, an implicit linear structure in Camassa-Holm equation
with dispersion is found. According to the linear structure, we give
the superposition of multi-solutions to Camassa-Holm equation with
dispersion.
This is an interesting result. \\

\section{Classification of  traveling wave
solutions and superposition of multi-solution to Camassa-Holm
equation with dispersion}

Under the traveling wave transformation $u=u(\xi),
 \xi=x-ct$, Camassa-Holm equation with
dispersion is reduced to the following ODE
\begin{equation}\label{od}
(2k-c)u'+3uu'-\varepsilon(uu'''+2u'u''-cu''')+\gamma u'''=0,
\end{equation}
integrating once yields the following equation
\begin{eqnarray}\label{ODE}
u''+\frac{1}{2(u-c-\frac{\gamma}{\varepsilon})}(u')^2-
\frac{3u^2-2(c-2k)u+c_0}{2\varepsilon(u-c-\frac{\gamma}{\varepsilon})}=0,
\end{eqnarray}
where $c_0$ is an arbitrary constant.

So we only need to solve the Eq.(\ref{ODE}). We give the general
solution of
the Eq.(\ref{ODE}) in the following:\\

\textbf{Lemma}: The general solution of the Eq.(\ref{ODE}) is as
follows:
\begin{equation}
\pm(\xi-\xi_0)=\int
\sqrt{\frac{\varepsilon(u-c-\frac{\gamma}{\varepsilon})}{(u-c-\frac{\gamma}{\varepsilon})^3+
d_2(u-c-\frac{\gamma}{\varepsilon})^2+d_1(u-c-\frac{\gamma}{\varepsilon})+d_0}}\mathrm{d}u,
\end{equation}
where $\xi_0$and $d_0$  are two arbitrary constants, and
$d_2=2c+2k+\frac{3\gamma}{\varepsilon},
d_1=3(c+\frac{\gamma}{\varepsilon})^2+2(2k-c)(c+\frac{\gamma}{\varepsilon})+c_0$.

According to the above lemma, we give the classification of all
single traveling wave solutions to Camassa-Holm equation with
dispersion. We have the following result. \\

\textbf{Theorem 1}: All traveling wave solutions to Camassa-Holm
equation with dispersion can be
classified as follows:\\

Case 1: $\epsilon>0$.\\

Case 1.1. $d_0=0$. Denote $\triangle=d_2^2-4d_1$. There are two
cases to be discussed:\\

Case 1.1.1: If $\triangle=0$, then the corresponding solutions to
the Eq.(\ref{od}) are
\begin{equation}
u=c_1\exp(\pm\frac{\xi}{\sqrt
\varepsilon})-k-\frac{\gamma}{2\varepsilon},
\end{equation}
where $c_1$ is an arbitrary constant.

Case 1.1.2: If $\triangle>0$ or $\bigtriangleup<0$, then the
corresponding solutions of the Eq.(\ref{od}) are
\begin{eqnarray}
u=c_1\exp(\frac{\xi}{\sqrt \varepsilon})+c_2\exp(-\frac{\xi}{\sqrt
\varepsilon})-k-\frac{\gamma}{2\varepsilon}.
\end{eqnarray}
where $c_1$ and $c_2$ are two arbitrary constants.\\

Case 1.2. $d_0\neq0$. Denote
\begin{equation}
\triangle=-27(\frac{2d_2^3}{27}+d_0-\frac{d_1d_2}{3})^2-4(d_1-\frac{d_2^2}{3})^3,
\end{equation}
\begin{equation}
D_1=d_1-\frac{d_2^2}{3},
\end{equation}\\
where $ \triangle $ and $ D_1 $ make up a complete discrimination
system for $ F(u)=(u+ \frac{\omega}{k})^3+d_2(u+
\frac{\omega}{k})^2+d_1(u+ \frac{\omega}{k})+d_0 $.
There are the following four cases to be discussed:\\

Case 1.2.1: If $ \triangle=0, D_1<0 $, then we have $
F(u)=(u-\alpha)^2(u-\beta), \alpha\neq\beta. $ We take the change of
the variable as follows:
\begin{equation}u=\frac{\beta
v^2-c-\frac{\gamma}{\varepsilon}}{v^2-1},
\end{equation}
its inverse transformation is
\begin{equation}
\frac{u-c-\frac{\gamma}{\varepsilon}}{u-\beta}=v^2.
\end{equation}
 then we have
\begin{equation}
\pm\frac{1}{\sqrt
\varepsilon}(\xi-\xi_0)=\ln|\frac{v+1}{v-1}|+\sqrt{\frac{\alpha-c-\frac{\gamma}{\varepsilon}
}{\alpha-\beta}}\ln|\frac{v-\sqrt{\frac{\alpha-c-\frac{\gamma}{\varepsilon}}{\alpha-\beta}}}
{v+\sqrt{\frac{\alpha-c-\frac{\gamma}{\varepsilon}}{\alpha-\beta}}}|;
(\frac{\alpha-c-\frac{\gamma}{\varepsilon}}{\alpha-\beta}>0).
\end{equation}
and
\begin{equation}
\pm\frac{1}{\sqrt \varepsilon}(\xi-\xi_0)=\ln|\frac{v
+1}{v-1}|-2\sqrt{\frac{\alpha-c-\frac{\gamma}{\varepsilon}}{\beta-\alpha}}\arctan
(v\sqrt{\frac{\beta-\alpha}{\alpha-c-\frac{\gamma}{\varepsilon}}}),
(\frac{\alpha-c-\frac{\gamma}{\varepsilon}}{\beta-\alpha}>0).
\end{equation}
\\

Case 1.2.2: If $ \triangle=0, \ D_1=0 $, then we have $
F(u)=(u-\alpha)^3 $, the solution is as follows:
\begin{equation}
\pm\frac{1}{2\sqrt
\varepsilon}(\xi-\xi_0)=\pm\sqrt{\frac{u-c-\frac{\gamma}{\varepsilon}}{u-\alpha}}+
\frac{1}{2}\ln|\frac{\sqrt{\frac{u-c-\frac{\gamma}{\varepsilon}}{u-\alpha}}\mp1}
{\sqrt{\frac{u-c-\frac{\gamma}{\varepsilon}}{u-\alpha}}\pm1}|.
\end{equation}\\

Case 1.2.3: If $ \triangle>0, \ D_1<0 $, then $
F(w)=(w-\alpha)(w-\beta)(w-\rho) $, we suppose that $
\alpha>\beta>\rho $.  Take the change of variable
\begin{equation}u=\frac{\alpha
v^2-c-\frac{\gamma}{\varepsilon}}{v^2-1},
\end{equation}
then the corresponding integral becomes
\begin{eqnarray}
\pm\frac{\sqrt{(\alpha-\beta)(\alpha-\rho)}}{2\sqrt
\varepsilon(\alpha-c-\frac{\gamma}{\varepsilon})}(\xi-\xi_0)=\cr\int\{1+\frac{1}{2}(\frac{1}{v-1}
-\frac{1}{v+1})\}\frac{1}{\sqrt{(v^2+A)(v^2+B)}}\mathrm{d}v.
\end{eqnarray}
where $A=\frac{\beta-c-\frac{\gamma}{\varepsilon}}{\alpha-\beta},
B=\frac{\rho-c-\frac{\gamma}{\varepsilon}}{\alpha-\rho}$. It is easy
to see that the integral (35) can be expressed by the first kind of
elliptic integrals and the third kind of elliptic
integrals.\\

Case 1.2.4: If $ \triangle<0 $, then we have $
F(u)=(u-\alpha)(u^2+pu+q), \ \ p^2-4q<0. $  We take the change of
variable $ v=\frac{u-c-\frac{\gamma}{\varepsilon}}{u-\alpha}$, the
corresponding integral becomes
\begin{eqnarray}
\pm\frac{\xi-\xi_0}{\sqrt
\varepsilon(\alpha-c-\frac{\gamma}{\varepsilon})}=\int(1+\frac{1}{v-1})\frac{1}{\sqrt{v(Av^2+Bv+C)}}\mathrm{d}v,
\end{eqnarray}
where $A=p\alpha+q+\alpha^2,
B=-(2\alpha+p)(c+\frac{\gamma}{\varepsilon})-p\alpha-2q,
C=(c+\frac{\gamma}{\varepsilon})^2+p(c+\frac{\gamma}{\varepsilon})+q$,
moreover $B^2-4AC<0$. It is easy to see that the corresponding
integral (36) can be expressed by the first kind of elliptic
integrals and the third
kind of elliptic integrals.\\

Case 2. $\varepsilon<0$.\\

Case 2.1. $d_0=0$. Denote $\triangle=d_2^2-4d_1$. If
$\triangle\leq0$, then there exists no solutions at all. So we only
consider the case of $\triangle>0$. When
$-2k-\frac{\gamma}{2\varepsilon}-\frac{\sqrt{\triangle}}{2}<u<-2k-
\frac{\gamma}{2\varepsilon}+\frac{\sqrt{\triangle}}{2}$, we have
\begin{equation}
u=-2k-\frac{\gamma}{\varepsilon}\pm\frac{\sqrt{\triangle}}{2}\sin(\frac{1}{\sqrt
{-\varepsilon}}(\xi-\xi_0)),
\end{equation}
or rewriting it by another form:
\begin{equation}
u=-2k-\frac{\gamma}{\varepsilon}+c_1\sin(\frac{1}{\sqrt
{-\varepsilon}}\xi)+c_2\cos(\frac{1}{\sqrt {-\varepsilon}}\xi),
\end{equation}
where $c_1$ and $c_2$ are two arbitrary constants. These are
periodic solutions.\\

Case 2.2. $d_0\neq0$.  $ \triangle, D_1 $ and $ F(u)$ are the same
with the former. There are also the following four cases to be discussed:\\

Case 2.2.1: If $ \triangle=0, D_1<0 $, then we have $
F(u)=(u-\alpha)^2(u-\beta), \alpha\neq\beta. $ We take the change of
the variable as follows:
\begin{equation}
u=\frac{\beta v^2+c+\frac{\gamma}{\varepsilon}}{1+v^2},
\end{equation}
its inverse transformation is
\begin{equation}
\frac{u-c-\frac{\gamma}{\varepsilon}}{u-\beta}=-v^2.
\end{equation}
 then we have
\begin{equation}
\pm\frac{1}{\sqrt {-\varepsilon}}(\xi-\xi_0)=2\arctan
v-\sqrt{\frac{\alpha-c-\frac{\gamma}{\varepsilon}
}{\beta-\alpha}}\ln|\frac{v-\sqrt{\frac{\alpha-c-\frac{\gamma}{\varepsilon}}{\beta-\alpha}}}
{v+\sqrt{\frac{\alpha-c-\frac{\gamma}{\varepsilon}}{\beta-\alpha}}}|;
(\frac{\alpha-c-\frac{\gamma}{\varepsilon}}{\beta-\alpha}>0).
\end{equation}
and
\begin{equation}
\pm\frac{1}{\sqrt {-\varepsilon}}(\xi-\xi_0)=2\arctan
v-2\sqrt{\frac{\alpha-c-\frac{\gamma}{\varepsilon}}{\alpha-\beta}}\arctan
(v\sqrt{\frac{\alpha-\beta}{\alpha-c-\frac{\gamma}{\varepsilon}}}),
(\frac{\alpha-c-\frac{\gamma}{\varepsilon}}{\alpha-\beta}>0).
\end{equation}\\

Case 2.2.2: If $ \triangle=0, \ D_1=0 $, then we have $
F(u)=(u-\alpha)^3 $, the solution is as follows:
\begin{equation}
\pm\frac{1}{2\sqrt
{-\varepsilon}}(\xi-\xi_0)=2\sqrt{\frac{u-c-\frac{\gamma}{\varepsilon}}{\alpha-u}}-
2\arctan(\sqrt{\frac{u-c-\frac{\gamma}{\varepsilon}}{\alpha-u}}).
\end{equation}\\

Case 2.2.3: If $ \triangle>0, \ D_1<0 $, then $
F(w)=(w-\alpha)(w-\beta)(w-\rho) $, we suppose that $
\alpha>\beta>\rho $.  Take the change of variable
\begin{equation}
u=\frac{\alpha v^2-c-\frac{\gamma}{\varepsilon}}{v^2-1},
\end{equation}
then the corresponding integral becomes
\begin{eqnarray}
\pm\frac{\sqrt{(\alpha-\beta)(\alpha-\rho)}}{2\sqrt
{-\varepsilon}(\alpha-c-\frac{\gamma}{\varepsilon})}(\xi-\xi_0)=\cr\int\{1+\frac{1}{2}(\frac{1}{v-1}
-\frac{1}{v+1})\}\frac{1}{\sqrt{-(v^2+A)(v^2+B)}}\mathrm{d}v.
\end{eqnarray}
where $A=\frac{\beta-c-\frac{\gamma}{\varepsilon}}{\alpha-\beta},
B=\frac{\rho-c-\frac{\gamma}{\varepsilon}}{\alpha-\rho}$. If $A>0$
and $B>0$, it is easy to see that no solution can be given. In other
cases, it is easy to see that the integral (28) can be expressed by
the first kind of elliptic integrals and the third kind of elliptic
integrals.\\

Case 2.2.4: If $ \triangle<0 $, then we have $
F(u)=(u-\alpha)(u^2+pu+q), \ \ p^2-4q<0. $  We take the change of
variable $ v=\frac{u-c-\frac{\gamma}{\varepsilon}}{\alpha-u}$, the
corresponding integral becomes
\begin{eqnarray}
\pm\frac{\xi-\xi_0}{\sqrt
{-\varepsilon}(\alpha-c-\frac{\gamma}{\varepsilon})}=\int(1-\frac{1}{v-1})\frac{1}{\sqrt{v(Av^2+Bv+C)}}\mathrm{d}v,
\end{eqnarray}
where $A=p\alpha+q+\alpha^2,
B=-(2\alpha+p)(c+\frac{\gamma}{\varepsilon})+p\alpha+2q,
C=(c+\frac{\gamma}{\varepsilon})^2-p(c+\frac{\gamma}{\varepsilon})+q$,
moreover $B^2-4AC<0$. It is easy to see that the corresponding
integral (29) can be expressed by the first kind of elliptic
integrals and the third
kind of elliptic integrals.\\

\section{An implicit linear structure and superposition of multi-solutions to CH-r equation}

According to the case 1.2 in Sec.2, we find an interesting fact,
that is, affine superposition of two solutions $ u_1=\exp(x-ct)$ and
$ u_2=\exp(-(x-ct))$ is also a solution to the Eq.(5). This is
because  the Eq.(5) has a special structure. In fact, the Eq.(5)
includes two parts, one part is $-cu'+c\varepsilon u'''$, another is
$2ku'+3uu'-\varepsilon uu'''-2\varepsilon u'u''+ru'''$. If we take
\begin{eqnarray}
-cu'+c\varepsilon u'''=0,\\
 2ku'+3uu'-\varepsilon
uu'''-2\varepsilon u'u''+ru'''=0,
\end{eqnarray}
then from the first equation, we have $u'=\varepsilon u'''$,
therefore the second equation becomes
$u+k+\frac{r}{2\varepsilon}=\varepsilon u''$. It is easy to see that
these two equations are compatible. Therefore, if we take
$v=u+k+\frac{r}{2\varepsilon}$, these two equations become linear
equations $v''=\frac{1}{\varepsilon}v$ essentially. In this case,
the superposition property of the solutions $v$ is found. Thus there
is a two dimensional linear subspace in the set of the solutions of
$v$-equations. When $\varepsilon>0$, the corresponding bases are
$e_1=\exp(x-ct)$ and $e_2=\exp(-(x-ct))$. When $\varepsilon<0$, the
bases are $e_1=\sin(\frac{1}{\sqrt{-\varepsilon}}\xi)$,
$e_2=\cos(\frac{1}{\sqrt{-\varepsilon}}\xi)$. This means that there
is a two dimensional affine subspace in the set of the
solutions of $u$-equations.\\

More general, we find an implicit linear structure in Camassa-Holm
equation with dispersion. In fact, we rewrite Camassa-Holm equation
with dispersion by
\begin{eqnarray}\label{GCH}
u_t-\varepsilon u_{xxt}=\varepsilon(2u_xu_{xx}+uu_{xxx})-\gamma
u_{xxx}-2ku_x-3uu_x,
\end{eqnarray}
letting the left and the right sides both are zeros yield two
equations
\begin{eqnarray}\label{GCH}
u_t-\varepsilon u_{xxt}=0,\\
\varepsilon(2u_xu_{xx}+uu_{xxx})-\gamma u_{xxx}-2ku_x-3uu_x=0.
\end{eqnarray}
Integrating the above first equation, we have
\begin{equation}
u=\varepsilon u_{xx}+f(t),
\end{equation}
thus $u_x=\varepsilon u_{xxx}$, if we take
$f(t)=-k-\frac{\gamma}{2\varepsilon}$, then the above second
equation is satisfied automatically. Therefore, we find a linear
structure in Camassa-Holm equation with dispersion, that is,
\begin{equation}
u+k+\frac{\gamma}{2\varepsilon}=\varepsilon u_{xx}.
\end{equation}
Let $v=u+k+\frac{\gamma}{2\varepsilon}$, then the above equation
becomes
\begin{equation}
v=\varepsilon v_{xx},
\end{equation}
when $\varepsilon>0$, its general solution is as follows
\begin{equation}
v(x,t)=\sum [g_i(t)\exp(\frac{x}{\sqrt
\varepsilon}-s_i(t))+h_i(t)\exp(-\frac{x}{\sqrt
\varepsilon}-r_i(t))],
\end{equation}
where $g_i(t), h_i(t), s_i(t)$ and $r_i(t)$ all are arbitrary
functions, thereafter. The corresponding $u$-solution is
\begin{equation}
u(x,t)=\sum [g_i(t)\exp(\frac{x}{\sqrt
\varepsilon}-s_i(t))+h_i(t)\exp(-\frac{x}{\sqrt
\varepsilon}-r_i(t))]-k-\frac{\gamma}{2\varepsilon}.
\end{equation}
When $\varepsilon<0$, the general $v$-solution is
\begin{equation}
v(x,t)=\sum [g_i(t)\sin(\frac{x}{\sqrt
{-\varepsilon}}-s_i(t))+h_i(t)\cos(-\frac{x}{\sqrt
{-\varepsilon}}-r_i(t))],
\end{equation}
the corresponding $u$-solution is
\begin{equation}
u(x,t)=\sum [g_i(t)\sin(\frac{x}{\sqrt
{-\varepsilon}}-s_i(t))+h_i(t)\cos(-\frac{x}{\sqrt
{-\varepsilon}}-r_i(t))]-k-\frac{\gamma}{2\varepsilon}.
\end{equation}
Through above illustration, we show easily the affine superposition
of multi-solutions to Camassa-Holm equation with dispersion using
implicit
linear structure. In summary, we have the following conclusions.\\

\textbf{Theorem 2}: When $\varepsilon>0$, Camassa-Holm equation with
dispersion has solution as follows:
\begin{equation}
u(x,t)=\sum [g_i(t)\exp(\frac{x}{\sqrt
\varepsilon}-s_i(t))+h_i(t)\exp(-\frac{x}{\sqrt
\varepsilon}-r_i(t))]-k-\frac{\gamma}{2\varepsilon};
\end{equation}
when $\varepsilon<0$, the corresponding solution is
\begin{equation}
u(x,t)=\sum [g_i(t)\sin(\frac{x}{\sqrt
{-\varepsilon}}-s_i(t))+h_i(t)\cos(-\frac{x}{\sqrt
{-\varepsilon}}-r_i(t))]-k-\frac{\gamma}{2\varepsilon}.
\end{equation}
where $g_i(t), h_i(t), s_i(t)$ and $r_i(t)$ all are arbitrary
functions.

\section{Conclusions}

In summary, according to the invariance property of the Eq.(5) under
the one-parameter group, we reduce it to the first order ODE and
give its general solutions. Furthermore applying complete
discrimination system for polynomial, we obtain the classifications
of all single traveling wave solutions to Camassa-Holm equation with
dispersion. Those complete results can't be given by any other
indirect methods. In addition, we find that a two dimensional affine
subspace structure in the set of solutions of the reduced ODE can be
obtained. More general, we find an implicit linear structure. Using
the linear structure, we give easily the affine superposition of
multi-solutions to Camassa-Holm equation with
dispersion.\\

 \textbf{Remark}: Because Camassa-Holm equation with
dispersion becomes CH equation as
 $\varepsilon=1, r=0$, so we will give the corresponding
 results to CH equation in this parameters case.\\

 \textbf{Acknowledgements}: Thanks to professor D.D.Holm for his suggestions.


\begin{thebibliography}{2}
\bibitem{CH}R Camassa and D D Holm:  \emph{Phys. Rev. Lett.} 71(1993) 1661.
\bibitem{CHH}R Camassa, D D Holm and J M Hyman: \emph{Adv. Appl. Mech.} 31(1994)1.
\bibitem{ACH}M S Alber, R Camassa , D D Holm and J E Marsden: \emph{Lett. Math. Phys.}
32(1994) 137.
\bibitem{ACH2}M S Alber, R Camassa , D D Holm and J E Marsden:   \emph{Proc. R. Soc. Lond. A}
450(1996) 677.
\bibitem{shi}J Schiff: \emph{Phys. D} 121(1998) No1-2 24.
\bibitem{Mc}H Mvkeam:  \emph{Comm. Pure. Appl. Math.} 57(2004) 416.
\bibitem{Joh}R S Johnson: R.Soc. Lond. Proc. Ser. A Math.
Phys. Eng. Sci. 459(2003)1687.
\bibitem{FOC}A S Focas:  \emph{Acta. Appl. Math.} 39(1995) 295.
\bibitem{LO}Y A Li and P J olver: \emph{Discr. Cont. Dyn. Sys.} 3(1997)419.

\bibitem{Ros1}P Rosenau: \emph{Phys.Rev. Lett.} 73(1994) 1737.
\bibitem{Ros2}P Rosenau:\emph{Phys. Lett.A} 211 (1996) 265.
\bibitem{Ros3}P Rosenau and J M Hyman: \emph{hys.Rev. Lett.} 70(1993)564.
\bibitem{lz1}Z R Liu R Q Wang and Z J Jing: Chaos, Solitons and
Fractals, 19(2004)77.
\bibitem{lz2}Z R Liu and T F Qian:Int. Jour. of Bif. and chaos.
11(2001)781
\bibitem{lz3}Z R Liu and T F Qian:Appl. Math. Model.26(2002)473.
\bibitem{d1}H R Dullin:phys.Rev.Lett. 87(2001)194501.
\bibitem{qi}T F Qian and M Y Tang:Chaos, Solitons and Fractals,
12(2001)1347.
\bibitem{bo}J P Boyd:Appl. Math. Comput. 81(1997)173.
\bibitem{guo}B L Guo and Z R Liu:Science in China A 46(2003)696.
\bibitem{zh}W L Zhang:Science in China A 47(2004)862.
\bibitem{ta}M Y Tang and C X Yang:Chaos, Solitons and Fractals.20(2004)815.

\end{thebibliography}
\end{document}